# Molecular-based coordination polymer as reversible and precise acetonitrile electro-optical readout

Esther Resines-Urien,[a] Enrique Burzurí,[a*] Estefania Fernandez-Bartolome,[a] Miguel Ángel García García-Tuñón,[b] Patricia de la Presa,[c] Roberta Poloni,[d] Simon J. Teat,[e] and Jose Sanchez Costa[a*]

[a]IMDEA Nanociencia, C/ Faraday 9, Ciudad Universitaria de Cantoblanco, 28049, Madrid, Spain.
[b]Instituto de Cerámica y Vidrio, CSIC, C/Kelsen s/n, 28240 Madrid, Spain.
[c]Instituto de Magnetismo Aplicado, UCM-ADIF-CSIC, A6 22.500 km, 28230 Las Rozas, Spain.
[d]Université Grenoble Alpes, CNRS, SIMAP, 38000 Grenoble, France
[e]Advanced Light Source (ALS), Berkeley Laboratory, 1 Cyclotron Road, Berkeley, CA 94720, USA.

**Efficient detection of harmful organic volatile compounds is a major health and environmental need in industrialized societies. Tailor-made metal-organic frameworks and other coordination polymers are emerging as promising sensing molecular materials thanks to their responsivity to a wide variety of external stimuli, and could be used to complement conventional sensors. Here, a non-porous crystalline 1D Fe(II) coordination polymer acting as a porous acetonitrile host is presented. The desorption of interstitial acetonitrile is accompanied by magneto-structural transitions easily detectable in the optical and electronic properties of the material. The structural switch and therefore its (opto)electronic readout are reversible under exposition of the crystal to acetonitrile vapor. Coordination polymers can be versatile sensors for volatile acetonitrile and potentially other organic compounds.**

Atmospheric and water pollution has become the preeminent global problem of our age.[1] The increasing emissions of volatile organic compounds (VOCs) and their resulting impact on air quality is one of these major environmental concerns.[2] Some VOCs are identified as highly toxic or carcinogenic and may cause impact on human health as well as on the natural ecosystem.[2] VOCs are emitted from the use of many everyday household products which makes the control of their emissions particularly difficult and critical. Consequently, a number of major environmental safety agencies have established guidelines to limit the exposure of humans.[2] Highly sensitive analytical techniques are currently employed for the accurate quantification of VOCs. However, these techniques display some drawbacks, such as their low portability, constrained selectivity and high cost.[2,3]

To overcome these limitations porous metal-organic frameworks (MOFs) have attracted a great deal of attention. Briefly, a MOF is a crystalline solid made by the self-assembly of single metal cations or metal clusters together with organic bridging ligands[2,4] with potential applications in catalysis, magnetism, electronics, luminescence, gas storage and drug delivery.[3,5–7] In addition, these materials are found to be excellent candidates for sensing applications. Various external stimuli, such as guest molecule adsorption-desorption, can modify the physical properties and dynamic behavior of the MOF,[2,8,9] opening the possibility to their application as switches and molecular sensors.[2,6,8,9] While usually porous solids have been mainly exploited for these applications, less attention has been paid to non-porous materials showing the ability to incorporate VOCs by internal structural reorganization, probably because of the lack of structural data.[6,9–12] Nevertheless, both approaches, porous and non-porous materials can uptake molecules, producing an easy-to-measure response.[6] Ideally, these chemosensors should show easily detectable changes in physicochemical properties, such as changes in the color of the material,[2,6] the luminescent emission,[3,7] the electrical conductivity[2,6] or the magnetic behavior.[9]

Here we present a novel non-porous crystalline 1D Fe(II) coordination polymer acting as porous material hosting acetonitrile. We show that the release of interstitial acetonitrile from the crystal under increasing temperature is accompanied by sharp magneto-structural transitions that can be detected as a change of crystal color, from yellow to orange, and as an abrupt high-current resonance in the electron transport through the crystal. Interestingly, the initial structure and therefore color and electrical transition are recovered once the crystal is exposed again to acetonitrile. This coordination polymer can be therefore used as a reversible and precise (opto)electronic detector for accurately monitoring ambient acetonitrile and potentially other VOCs.

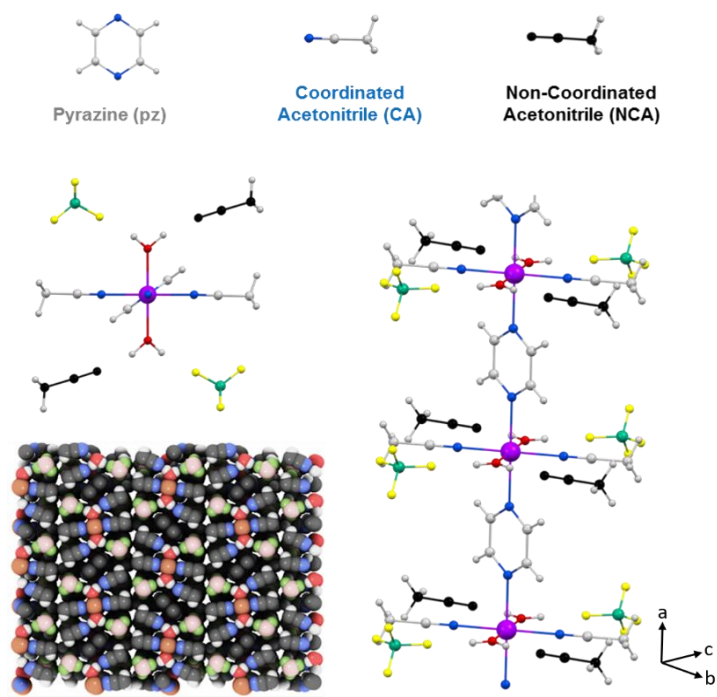

**Figure 1.** Crystal structure of **1**·2CH$_3$CN, where Fe is represented in purple, C in grey, N in blue, O in red, B in green, F in yellow, H in white and the uncoordinated acetonitrile in black.

The $_\infty${[Fe(H$_2$O)$_2$(CH$_3$CN)$_2$(pyrazine)](BF$_4$)$_2$·(CH$_3$CN)$_2$} complex, hereafter (**1**·2CH$_3$CN), was synthesized from a reaction of pyrazine with Fe(BF$_4$)$_2$·6H$_2$O carried out in acetonitrile (see Supporting Information). After one day, yellow needle-shaped crystals appeared, which crystallized in the orthorhombic space group *Cmca* (Table S1). In this structure there is one crystallographically independent iron (II), which is coordinated octahedrally with two H$_2$O molecules, two acetonitriles (CH$_3$CN) and two pyrazines. Pyrazine acts as a bridging ligand between two neighboring irons, resulting in the formation of 1D polymeric chains along the *a* axis (Figure 1). The distance between Fe-N (pyrazine) is 2.225 Å, and Fe-N (CH$_3$CN), 2.163 Å. These distances are characteristic of high spin (HS) iron (II).[13,14]

There is no direct inter-chain interaction although they interact *via* the BF$_4^-$ counterion, which forms a hydrogen bond with a water molecule of one chain, with F-H (H$_2$O) bond lengths of 1.795 Å (F-O: 2.655 Å), and a Van der Waals interaction with a coordinated acetonitrile of the adjacent chain, with F-H (CH$_3$CN) of 2.270 Å (F-C: 3.066 Å). These interactions occur along the *b* axis, but there is no interaction along the *c* axis, as can be seen in Figure 1. In addition, the anion also interacts with the pyrazine ring, where F-H (pyrazine) bond lengths are equal to 2.657 Å (F-C: 3.276 Å). The interstitial acetonitrile forms a hydrogen bond with the water molecule, with a distance O-H of 1.862 Å (O-N: 2.744 Å), and a BF$_4^-$ counterion, F-H 2.667 Å (F-C: 3.193 Å).

Figure 2a shows the optical reflectivity (OR) of a **1**·2CH$_3$CN crystal measured while heating it from 288 K up to 373 K. Initially, the OR remains constant and no change in the color of the crystal is observed as seen in Figure S1. Interestingly, the OR sharply increases at around 305 K and the **1**·2CH$_3$CN crystal changes color to a brighter shade of yellow, as seen in Figure S2. This change is also clearly seen in the video S1 in the Supplemental Material. The OR remains approximately constant from 306 K up to 316 K and thereafter gradually decreases (or sharply depending on the sample) as the color of the crystal changes from yellow to orange (see Figure 2b). At 355 K, **1**·2CH$_3$CN crystal is completely orange and the OR stabilizes to a constant value. In addition, a decrease in the volume of the crystal can be observed (Videos S2 and S3). Upon cooling down back to room temperature no change in color was observed, suggesting that the change in the optical reflectivity is not temperature reversible.

The infrared IR spectrum of **1**·2CH$_3$CN is measured in the same range of temperatures to determine whether the changes in the OR are related to structural changes in the molecules. The low energies spectrum is shown in Figure 2c. See Figures S3 to S6 for the full energy scale spectrum and a detailed analysis. The room temperature spectrum shows the bands associated to the different constituents of the polymer: acetonitrile, coordinated water molecules, pyrazine and the broad band of the BF$_4^-$ anions at 1016 cm$^{-1}$. Interestingly, a gradual disappearance of the band at 766 cm$^{-1}$ is clearly observed as the temperature increases. The same behavior is observed for the bands around 2250 cm$^{-1}$ (see Fig. S3-S6). The vanishing bands can be respectively assigned to the C≡N bend overtone[15,16] and the normal C≡N stretching mode that are present in the uncoordinated acetonitrile of the **1**·2CH$_3$CN. Their reduction and disappearance therefore suggests that an acetonitrile desolvation[17] from the crystal occurs as the temperature increases. On the other hand, the structural integrity of the crystal is conserved, which is an indicative that the acetonitrile groups coordinated to the Fe(II) remain in the structure. To further prove this, we have compared the **1**·2CH$_3$CN IR spectra with that of the analogue $^1_\infty${[Fe(CH$_3$CN)$_4$(pyrazine)](ClO$_4$)$_2$} with similar structure.[13] The difference is that no uncoordinated acetonitrile molecules are present in its lattice. The band at 766 cm$^{-1}$ does not appear in the spectrum at any temperature (Figure S7) and the band around 2250 cm$^{-1}$ is strongly reduced. Therefore, the presence and vanishing of the bands in **1**·2CH$_3$CN can be unambiguously assigned to the vanishing of uncoordinated acetonitrile with increasing temperature. The process is not temperature reversible and the IR bands do not reappear when cooling down to room temperature, in agreement with the OR measurements (Fig. S6). These results would be consistent with an irreversible loss of acetonitrile in the crystal structure.

To further explore the reversibility of the optical and structural transitions induced by the loss of acetonitrile in the crystal, a drop of acetonitrile is added to the "dry" orange crystal at room temperature after one temperature cycle. The drop is thereafter let dry in ambient conditions to eliminate any non-reabsorbed acetonitrile from the surface. Strikingly, color and optical reflectivity returns back to the initial yellowish characteristic as seen in Figure 2b and Figure S1. In addition, the absorption band at 766 cm$^{-1}$ reappears in the IR spectrum (Figure S8). The initial properties are also recovered when **1**·2CH$_3$CN is exposed to acetonitrile *via* vapor diffusion.

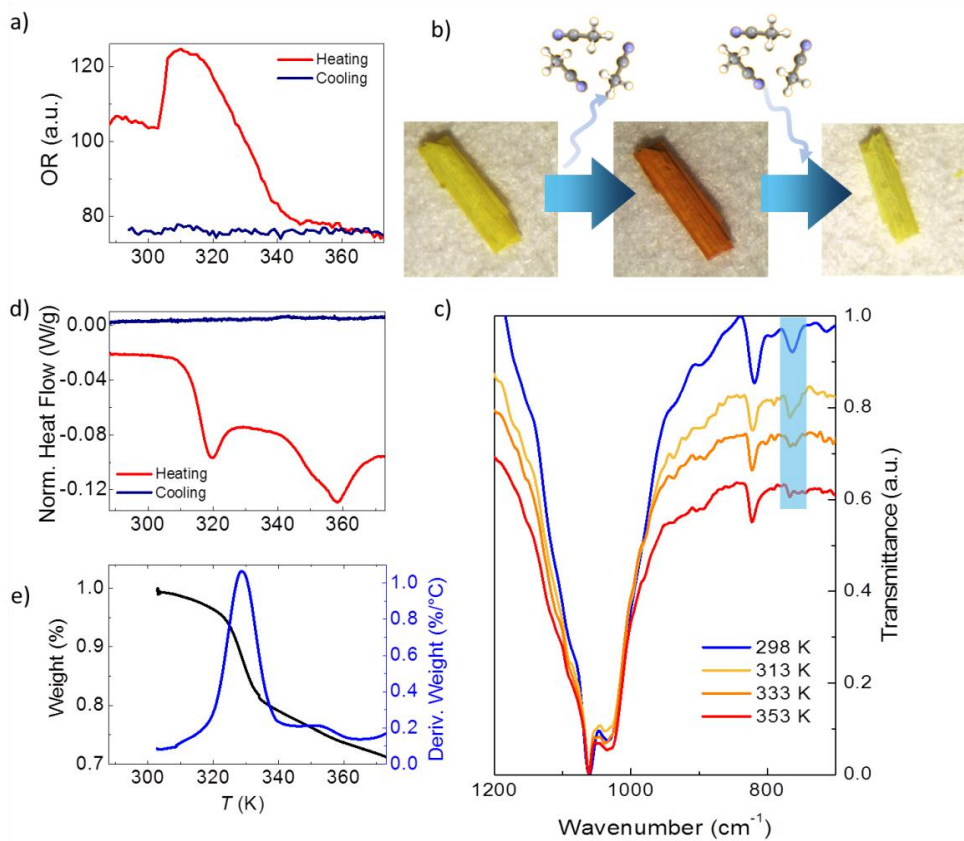

**Figure 2.** a) **1**·2CH$_3$CN optical reflectivity measurement between 288 K and 373 K. b) Optical image of a **1**·2CH$_3$CN crystal at room temperature (yellow), after heating and subsequent loss of acetonitrile molecules (orange) and after exposition to acetonitrile (yellow). c) **1**·2CH$_3$CN infrared spectra between 1200 and 700 cm$^{-1}$ at different temperatures. d) DSC measurements on a **1**·2CH$_3$CN crystal between 288 K and 373 K. e) TGA curve measured between 303 K and 373 K (black line). The blue line represents the derivative weight.

Further proof of the structural origin behind the color and OR transitions is provided by the differential scanning calorimetry (DSC) measured between 288 K and 373 K. Two clear endothermic peaks, i.e. the crystal is absorbing energy, can be observed during the heating process (Figure 2d). The first one is centered around 315 K and is consistent with the first change in the optical reflectivity. The second peak is centered at 358 K and occurs at the same temperature as the OR becomes lowest and constant and the crystal has completely changed from yellow to dark orange (see Figure S1). The endothermic nature of the peaks is indicative of a crystalline phase transition involving breaking of molecular bonds or the melting/evaporation of molecular species present in the crystal. No peaks, either endothermic or exothermic, were observed when cooling the sample from 373 K back to 288 K, as observed also in the OR. Interestingly, both structural transitions are also translated into magnetic anomalies in the magnetic susceptiblity at the same temperatures (see discussion in the Supporting Information).

The thermogravimetric analysis (TGA) of **1**·2CH$_3$CN was measured between room temperature and 873 K (see Figure 2e and S9). A first weight loss triggers at around 315 K (first transition in OR and DSC) and finishes at 345 K (the OR and color become constant). The total weight loss in this temperature range is 15.73 % and approximately fits with the loss of two acetonitrile molecules (16.08 %) per formula unit. This confirms that only the Fe-uncoordinated acetonitrile molecules evaporate from the structure at this temperature.

The effect of the structural transitions in the electronic properties of **1·2CH₃CN** has been studied in the same range of temperatures. Figure 3a shows the current $I$ measured across a single **1·2CH₃CN** crystal in nitrogen atmosphere at a fixed $V = 1$ V while ramping up the temperature from 288 K to 373 K (red curve) and consecutively down to the initial temperature (blue curve). Additional measurements on different crystals reproducing the same transport features can be found in Figure S10.

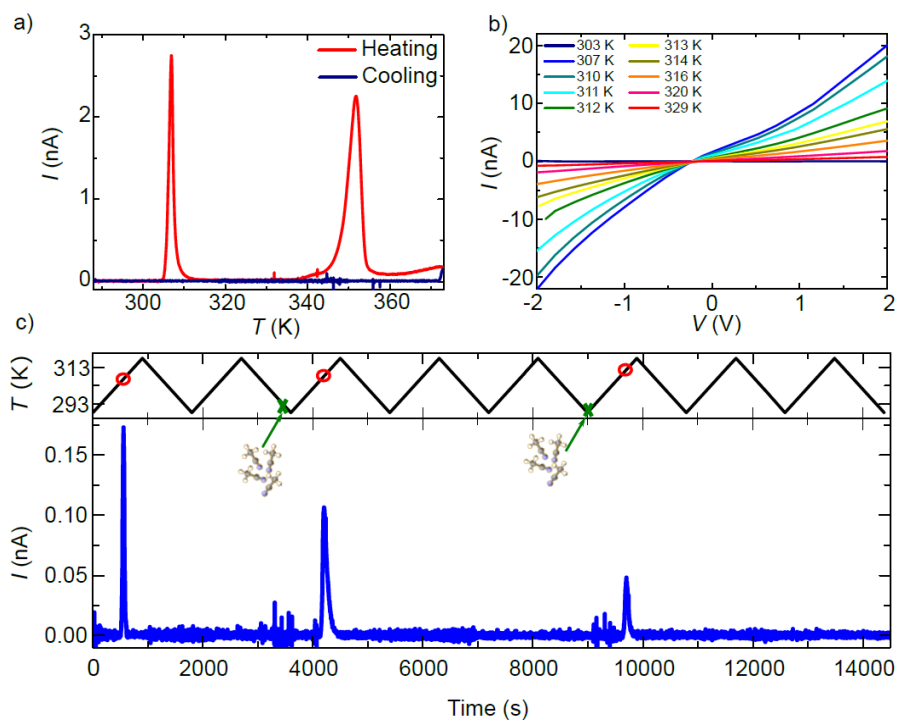

**Figure 3.** a) Electrical current $I$ vs. temperature $T$ measured in a single **1·2CH₃CN** crystal. Two sharp high-conductance peaks appear centered at 306 K and 353 K while increasing $T$. The peaks do not show up when the temperature is ramped down to the initial value. b) Current-voltage $IV$ characteristics measured at different temperatures around the first conductance peak. c) Proof-of-concept sensor of acetonitrile. A resonance in the current appears at a specific temperature (red circles) only in those temperature cycles where the **1·2CH₃CN** crystal is exposed to acetonitrile (green crosses).

Initially the crystal behaves almost as an insulator with current levels below $10^{-11}$ A, in agreement with the large band gap (~3 eV) computed by DFT for this polymer (see Supporting Information). As the temperature is increased, a sharp increment of the current of about 2 orders of magnitude sets on at 306 K followed by a softer decay back to low current levels. At higher temperatures, a second wider high-current peak appears between 343 and 353 K. Note that the magnitude and position of the peaks in temperature is reproducible in all the measured crystals (see Figure S10). Figure 3b shows the current-voltage $IV$ characteristics measured at different temperatures around the first peak in conductance. The current follows the same dependence in temperature observed in Figure 3a. The shape evolves from an insulating-like material to a clear non-linear dependence with bias, indicating the presence of Schottky barriers at the interface between electrodes and material or between grain boundaries in the material.

Interestingly, the current levels after the transition temperature remain higher than before the transition (dark blue curve at 303 K in Figure 3b).

A careful comparison with the structural and optical characterization of crystalline **1**·2CH$_3$CN shows that the high-current peaks match in temperature with the two respective anomalies in the OR and the calorimetric response. The sharp change in the conducting properties of the crystal is therefore univocally connected to the structural transitions undergone by the **1**·2CH$_3$CN crystal at around 306 K and 353 K.

None of the current peaks is present when the temperature is ramped back down to 288 K (blue curve in Figure 3a) and do not appear if the temperature is ramped up again, as also observed in the optical reflectivity and DSC. Strikingly, the two high-conductance peaks reappear in the charge transport measurement at identical temperatures after the **1**·2CH$_3$CN crystal is exposed to acetonitrile, as seen in Figure S11. The changes in the conductance are therefore reversible upon acetonitrile absorption-desorption in the **1**·2CH$_3$CN lattice. A simple conductance measurement can in principle be used as method to detect ambient acetonitrile. A proof-of-concept of such sensor is shown in Figure 3c. The current is measured across a single **1**·2CH$_3$CN crystal while cycling the temperature between 288 K and 318, i.e. around the first peak in conductance. A peak in the current shows up at roughly the same temperature (red circles) in those cycles where the crystal is exposed to acetonitrile vapor (green crosses), whereas it is absent in the rest of the cycles. A second example is shown in Figure SX.

The origin of the sharp change in conductance is intriguing since MOFs and coordination polymers are typically insulators or bad conductors[18,19] except a few remarkable cases[4,20–23]. A possible explanation for the first high-conductance peak can be connected to the structural change and the augmented mobility of the interstitial acetonitrile in the **1**·2CH$_3$CN crystals, proved by IR spectroscopy, DSC measurements and the clear change in volume of the crystals (see videos 2 and 3). Reordering of small molecules in porous materials is known to induce structural changes in the chain-like architecture that in turn may induce transitions in the magnetic behavior[24] and possibly the charge transport properties. On the other hand, the rapid drop in conductance under increasing temperature after the transition seems to rule out static changes in the electronic structure of the polymer, like a change in the band gap. This scenario is in agreement with our DFT calculations that show a sizeable change in the structure of the polymer after removal of the acetonitrile molecules, but a negligible change in the band gap, consistent with weakly interacting molecules (see binding energies and density of states in Figure S15).

An alternative but related explanation for dynamic changes in the conductance has been proposed for porous silicon and metal oxides detectors exposed to acetonitrile[25–27] and other polar solvents with high dielectric constants. The rise in current is not related to static changes in the bandgap. Instead, it can be explained in terms of dynamic changes in the dielectric constant of the material after infiltration of the acetonitrile that in turn modifies the charge distribution in the material. The consecutive rapid decrease of the conductance can be explained in this case as a dynamic capacitive effect over time or by the fast evaporation of acetonitrile due to its low vapor pressure. This effect has also been observed in MOFs[28–30] and CNT-polymer hybrid materials[19] This fast evaporation is clearly observed in the video 4? included in the Supporting Information. An increment of the surface brightness of the crystal is followed by the release of acetonitrile gas seen as blurriness in the image.

The second high-conductance peak seems also related to the second structural transition observed in the optical reflectivity and the DSC measurements. Interestingly, the peak shows up at exactly the acetonitrile boiling point temperature (355 K), indicating that the change in conductance is again linked to the reordering of acetonitrile.

To summarize we have presented a new non-porous crystalline one-dimensional coordination polymer acting as a porous material hosting acetonitrile. The material presents a reversible magneto-structural transition under desorption/absorption of acetonitrile molecules in the crystalline structure. The change in the structure is in turn translated into a distinct response in the electron transport through the material and its optical properties at well-defined temperatures close to ambient conditions. This family of non-porous materials can therefore be a versatile platform to fabricate tailor-made detectors for specific volatile organic compounds and with a versatile variety of read-out options from optical, magnetic to electron transport measurements. The 1D and nanoscale nature of the Fe coordination polymer backbone has the potential for organic molecule detectors with high spatial resolution.

**Acknowledgments**

JSC is grateful to the Spanish MINECO for financial support through National Research Project (CTQ2016-80635-P), the Ramon y Cajal Research program (RYC-2014-16866) and the Comunidad Autónoma de Madrid (PEJD-2017-PRE/IND-4037) for funding support. EB thanks funds from the MSCA-IF European Comission programme (grant 746579) and Programa de Atracción del Talento de la Comunidad de Madrid (2017-T1/IND-5562). IMDEA Nanociencia acknowledges support from the 'Severo Ochoa' Programme for Centres of Excellence in R&D (MINECO, Grant SEV-2016-0686). This research used resources of the Advanced Light Source, which is a DOE Office of Science User Facility under contract no. DE-AC02-05CH11231. We would like to thank to XALOC-ALBA synchrotron source under the project no. (2018012561). PP acknowledges financial support from the Spanish MINECO project MAT2015-67557-C2-1-P.

**Conflicts of interest**

There are no conflicts to declare.

# Molecular-based coordination polymer as reversible and precise acetonitrile electro-optical readout

## Supporting Information

Esther Resines-Urien,[a] Enrique Burzurí,[a*] Estefania Fernandez-Bartolome,[a] Miguel Ángel García García-Tuñón,[b] Patricia de la Presa,[b] Roberta Poloni[c], Simon J.Teat,[d] and José Sánchez Costa[a*]

1) Experimental Section.

2) Synthesis of $_\infty\{[Fe(H_2O)_2(CH_3CN)_2(pyrazine)](BF_4)_2\cdot(CH_3CN)_2\}$

3) Crystal Data, structural Refinement and selected distances and angles.

4) Crystal color variation. Reversibility under acetonitrile absorption/desorption.

5) Additional infrared absorption spectra.

6) 1·2CH$_3$CN thermogravimetric analysis.

7) Additional electron transport measurements. Reproducibility and reversibility.

8) Low temperature magnetic characterization.

9) DFT modelling.

# 1) Experimental Section

**Materials.** Chemicals and reagents were purchased from commercial suppliers and used as received.

**Physical measurements.**

- Crystal Structure Determination: The data were collected in a yellow needle crystal of **1**·2CH$_3$CN with a MD2M – Maatel diffractometer at the XALOC beamline (BL13) at ALBA Synchrotron with the collaboration of ALBA staff, from a Silicon (111) mono- chromator (T = 100 K, λ = 0.71073 Å).[32] The crystal was taken directly from its solution, mounted with a drop of Paratone-N oil and immediately put into the cold stream of dry N$_2$ on the goniometer. The structure was solved by direct methods and the refinement on $F^2$ and all further calculations were carried out with the SHELX-TL suite and OLEX2 program.[33]

- Optical reflectivity measurements between 288 and 373 K were performed using a MOTIC SMZ-171 optical stereoscope coupled with a MOTICAM 3. Images were collected in BMP format without any filter using the Motic Images Plus 3.0 software, with the mean value from each region of interest (ROI) analyzed under the ImageJ program. The temperature was controlled using a Linkam T95 system controller and a LNP 95 Liquid Nitrogen Cooling System.

- TGA was performed using a TA Instrument TGAQ500 with a ramp of 1 ºC/min under air from 303 to 873 K.

- FT-IR spectra were recorded as neat samples in the range 400-4000 cm$^{-1}$ on a Bruker Tensor 27 (ATR device) Spectrometer. The temperature dependence experiments were performed in the same instrument, using a thermometer and a heat gun.

- Elemental analyses (C, H and N) were performed on a LECO CHNS-932 Analyzer at the "Servicio Interdepartamental de Investigación (SIdI)" at Autónoma University of Madrid.

- Differential Scanning Calorimetry (DSC) measurements were performed in a TA Instruments Discovery MDSC 25 between 273 K and 393 K under a N$_2$ atmosphere with a ramp of 1°C/min. The sample was secured in a hermetically sealed aluminum sample pan.

- Magnetic susceptibility measurements between 10 K and 380 K were carried out in a Quantum Design MPMS-5S SQUID magnetometer under a 2000 Oe field. Each sample was secured inside a plastic capsule with cotton. Pascal constants were used to correct for the diamagnetic contribution.

- Electrical conductivity measurements were carried out on single crystals of **1**·2CH$_3$CN by a two-probe method using a Keithley 2450 SourceMeter under light and in a nitrogen atmosphere. Electrical contact to individual **1**·2CH$_3$CN crystals is made directly *via* the conducting tips of the electrical probe station. Charge transport is probed along the long axis of the crystals that coincides with the orientation of the molecular chains. The length of the crystals and the separation between the tips is typically a few hundred micrometers (200-500µm). The temperature was controlled using a Linkam T95 system with a LNP 95 Liquid Nitrogen Cooling System.

## 2)    Synthesis of $_\infty${[Fe(H$_2$O)$_2$(CH$_3$CN)$_2$(pyrazine)](BF$_4$)$_2$·(CH$_3$CN)$_2$}

This compound was synthetized at room temperature, dissolving 0.64 mmol of Fe(BF$_4$)$_2$·6H$_2$O in 2.5 ml of acetonitrile and adding it drop by drop to a solution of 1.2 mmol of pyrazine in 2.5 mL of acetonitrile. The resulting solution was stirred for 15 minutes and filtered. After one day, yellow crystals appeared.

Elemental analysis calculated (%) for **1**·2CH$_3$CN·0.6H$_2$O: C 27.69, H 4.10, N 16.14; found C 27.85, H 3.72, N 15.77.

FTIR (cm$^{-1}$): 3481, 2309, 2281, 2094, 1642, 1488, 1423, 1366, 1289, 1002, 817, 767, 589, 518, 466.

## 3)    Crystal Data and Structural Refinement and selected distances and angles

**Table S1.** Selected bond and interaction lengths [Å] for compound **1**·2CH$_3$CN.

| | |
|---|---|
| Fe1-N1 | 2.226(5) |
| Fe1-N2 | 2.164(5) |
| Fe1-O1 | 2.062(5) |
| F1-O1 | 2.655 |
| F1-H2W | 1.724 |
| F2-H1 | 2.816 |
| F2-C1 | 3.562 |
| F3-H3B | 2.546 |
| F3-C3 | 3.383 |
| N3-O1 | 2.742 |
| N3-H1W | 1.903 |
| N3-centroid(N1C1) | 3.232 |

**Table S2.** Crystallographic data of **1**·2CH$_3$CN.

| | |
|---|---|
| Compound | **1**·2CH$_3$CN |
| CCDC | |
| Chemical formula | C$_{12}$H$_{20}$B$_2$F$_8$FeN$_6$O$_2$ |
| Formula mass | 510.11 g/mol |
| Temperature (K) | 250.0 |
| Crystal system | Orthorhombic |
| Space group | *Cmca* |
| *a*/Å | 7.2430(17) |
| *b*/Å | 12.882(2) |
| *c*/Å | 23.908(6) |
| α/° | 90 |
| β/° | 90 |
| γ/° | 90 |
| *V*(Å$^3$) | 2230.8(9) |
| Z | 4 |
| Radiation type | Synchrotron |
| Density (calculated mg/m$^3$) | 1.518 |
| Absorption coefficient (mm$^{-1}$) | 0.76 |
| *F*(000) | 1032.0 |
| Crystal size (mm$^3$) | 0.000576 |
| Goodness of fit on *F*$^2$ | 1.112 |
| *R*1, *wR*2 [*I*>2σ(*I*)] | 0.0719, 0.2069 |
| *R*1, *wR*2 (all data) | 0.0930, 0.2330 |

## 4) Crystal color variation. Reversibility under acetonitrile absorption/desorption.

Figure S1 shows the color variation of a **1**·2CH$_3$CN crystal as the temperature is increased. The color changes from a pale yellow (298 K) to a dark orange (353 K) as the temperature is increased and interstitial acetonitrile is released. Strikingly, the crystal color turns back to its original pale yellow once it is exposed to acetonitrile vapor at room temperature (bottom panel). The optical properties are therefore reversible under acetonitrile absorption/desorption.

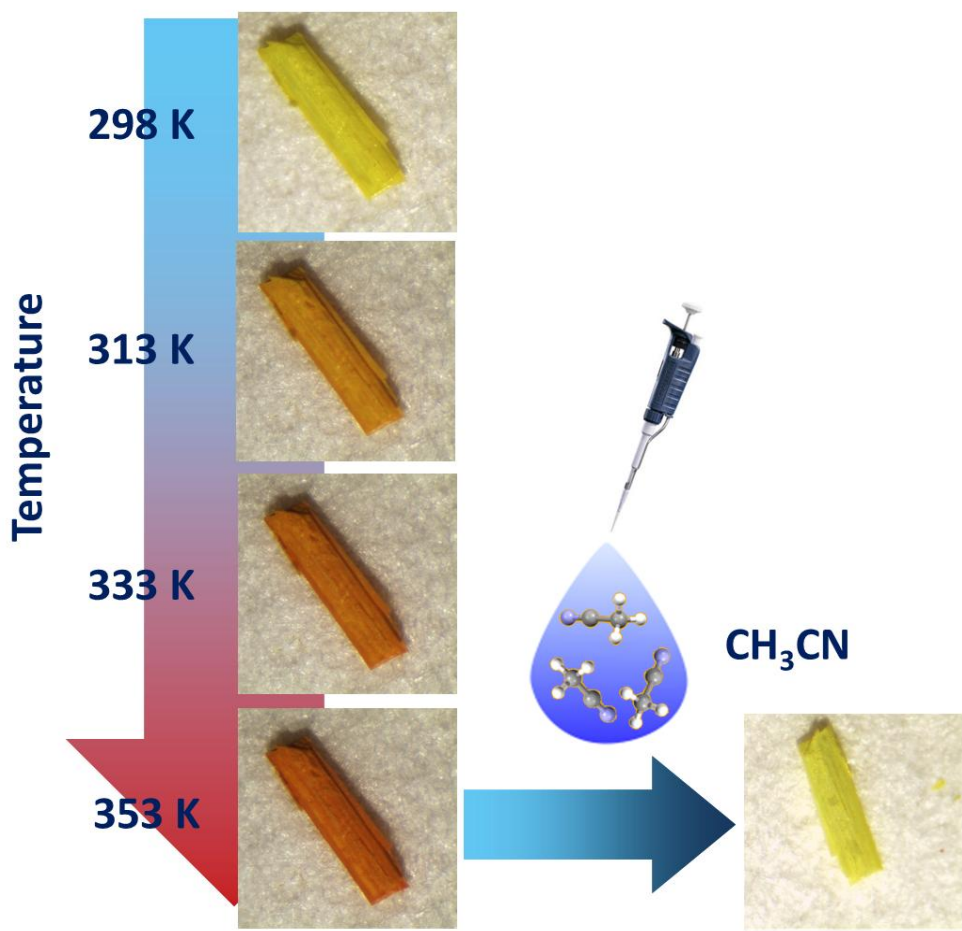

**Figure S1.** Color variation of a **1**·2CH$_3$CN crystal under increasing temperature. The color is recovered after a drop of acetonitrile is added at room temperature.

Figure S2 shows the color of a **1**·2CH$_3$CN crystal at three specific temperatures of special relevance in the OR and conductance measurements: a) before the first OR transition, b) at the first OR and conductance transition and c) at the second OR and conductance transition. At the first resonance (~310 K) a sharp change from a matte yellow to a shiny bright shade of yellow is observed. After that transition and coinciding with the progressive loss of acetonitrile observed in the IR and the TGA, the crystal color turns into darker orange.

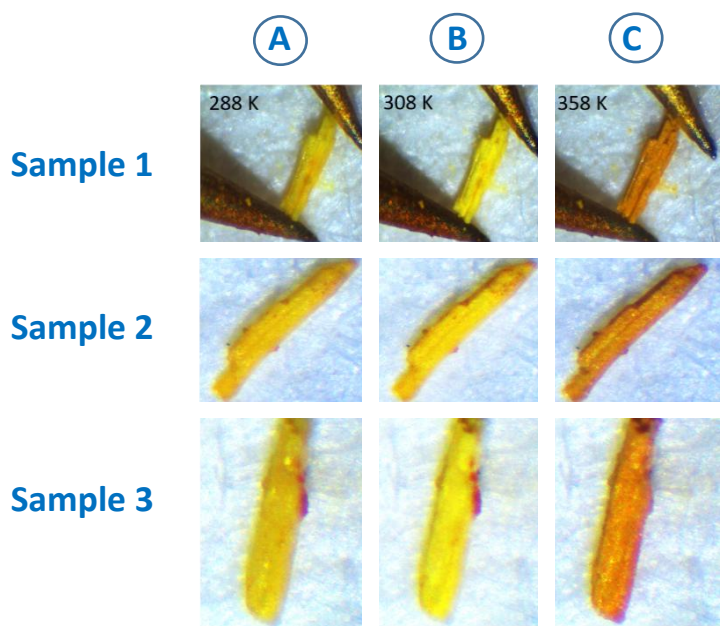

**Figure S2.** Color change in three different **1**·2CH$_3$CN crystals at three relevant temperatures for the OR and electron transport measurements.

## 5) Additional infrared spectra.

Figures S3 to S6 show the infrared absorption spectra of a **1**·2CH$_3$CN crystal at different temperatures. Figure S7 shows the room temperature infrared spectrum of the analogue compound pyrazine-ClO$_4$. Finally, Figure S8 shows the infrared spectrum of the **1**·2CH$_3$CN crystal shown in Figures S3-S6 after exposition to acetonitrile in ambient conditions.

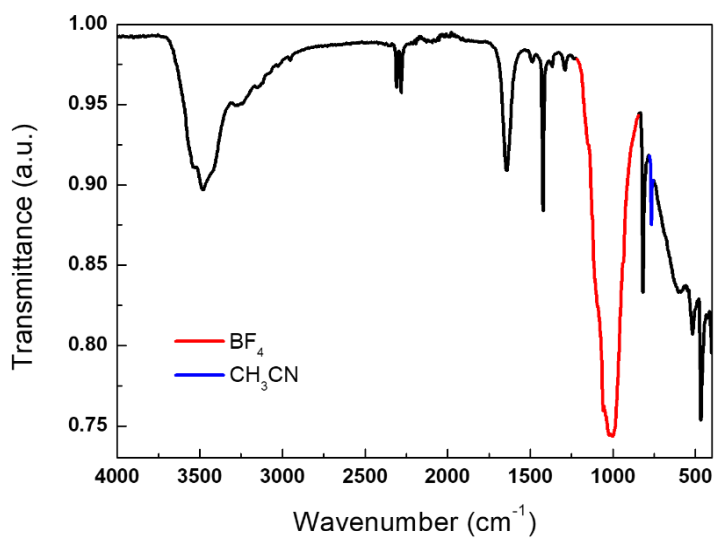

**Figure S3.** IR spectrum of **1**·2CH$_3$CN between 4000 cm$^{-1}$ and 400 cm$^{-1}$. The band corresponding with ν(BF$_4$) at 1022 cm$^{-1}$ is represented in red, while the C-C≡N bend overtone of acetonitrile, at 766 cm$^{-1}$, is colored in blue.

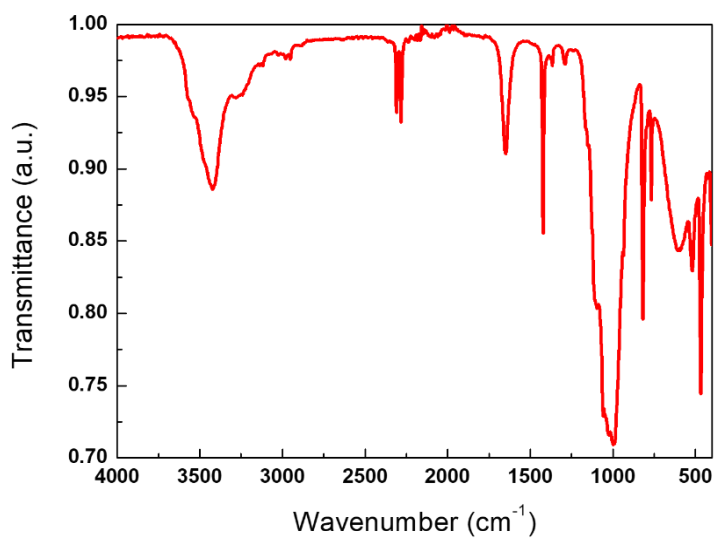

**Figure S4.** IR spectrum of **1**·2CH$_3$CN between 4000 cm$^{-1}$ and 400 cm$^{-1}$ at 317 K.

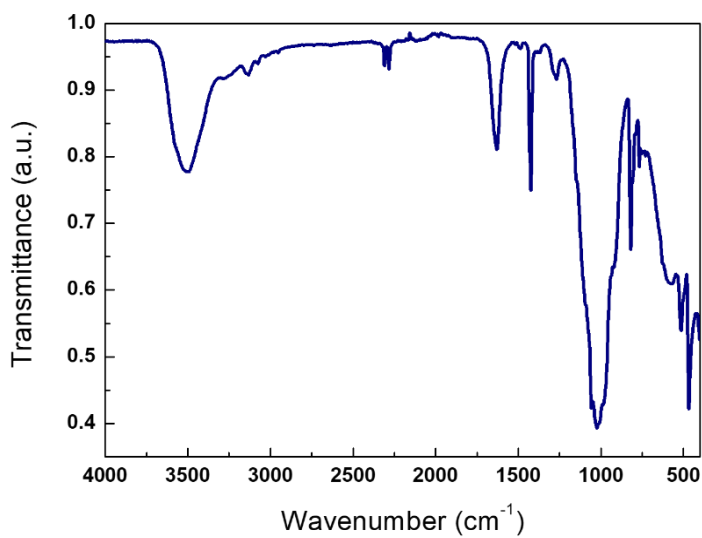

**Figure S5.** IR spectrum of **1**·2CH$_3$CN between 4000 cm$^{-1}$ and 400 cm$^{-1}$ at 333 K.

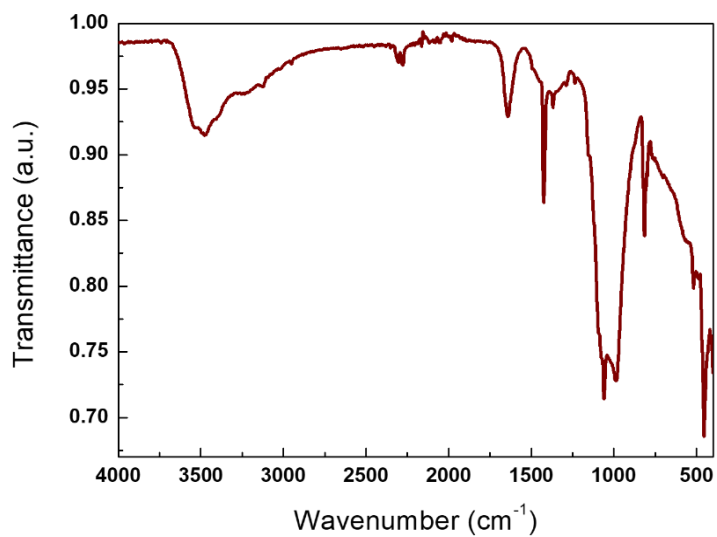

**Figure S6.** IR spectrum of **1**·2CH$_3$CN between 4000 cm$^{-1}$ and 400 cm$^{-1}$ at 353 K.

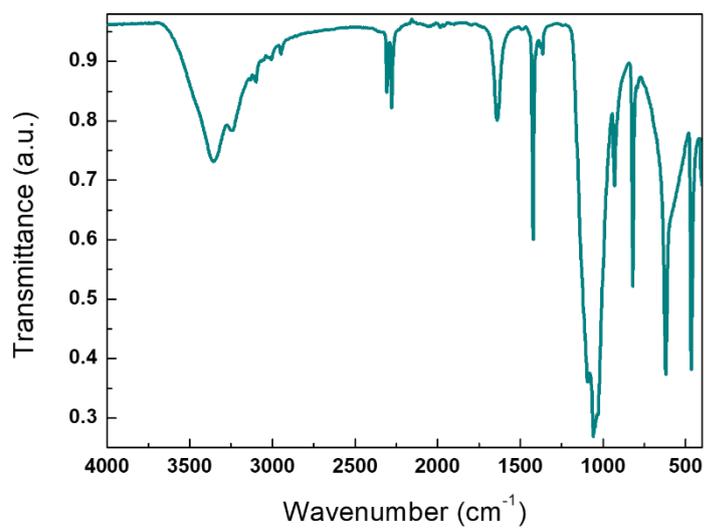

**Figure S7.** IR spectrum of pyrazine-ClO$_4$ between 4000 cm$^{-1}$ and 400 cm$^{-1}$.

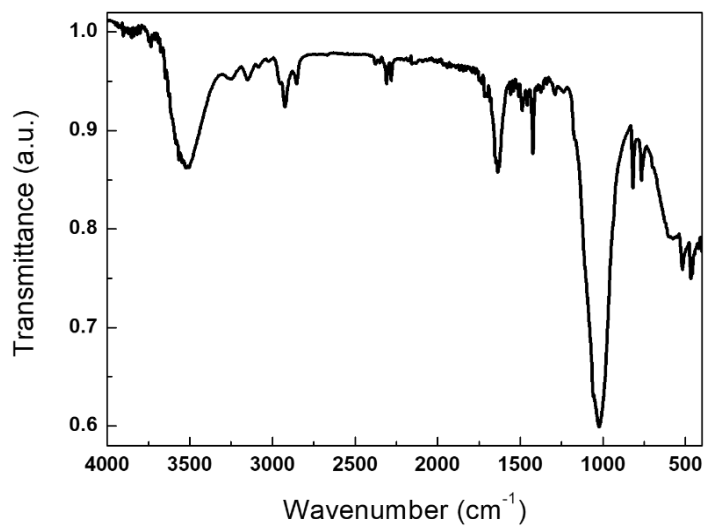

**Figure S8.** IR spectrum of **1**·2CH$_3$CN between 4000 cm$^{-1}$ and 400 cm$^{-1}$ at room temperature after a drop of acetonitrile was added to the "dry" crystal. The 766 cm$^{-1}$ band is recovered

## 6)  1·2CH$_3$CN thermogravimetric analysis

Thermogravimetric analysis extended to higher temperatures. It shows the peak corresponding to the loss of acetonitrile (see main text). The polymer decomposes completely above 440 K approximately.

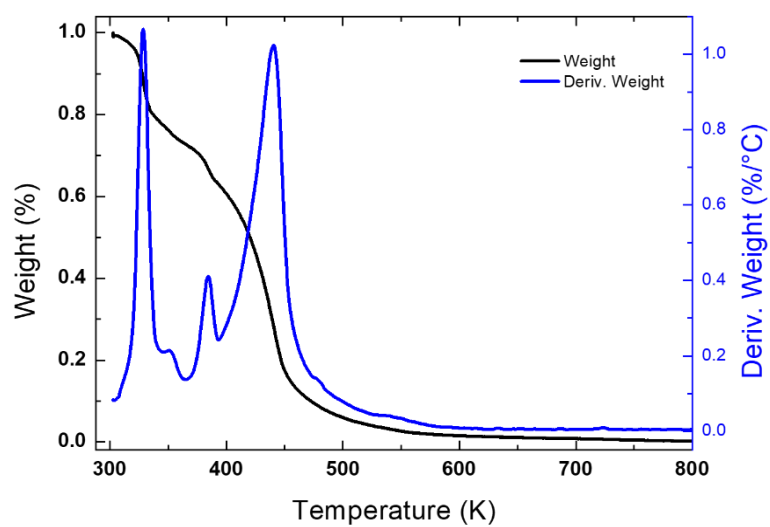

**Figure S9.** Thermogravimetric analysis of **1**·2CH$_3$CN between 303 K and 873 K.

# 7) Additional electron transport measurements. Reproducibility and reversibility.

Figure S10 shows the current as a function of the temperature measured in two additional **1**·2CH$_3$CN crystals. The magnitude and position of the conductance resonances are consistent with those shown in Figure 3 of the main manuscript.

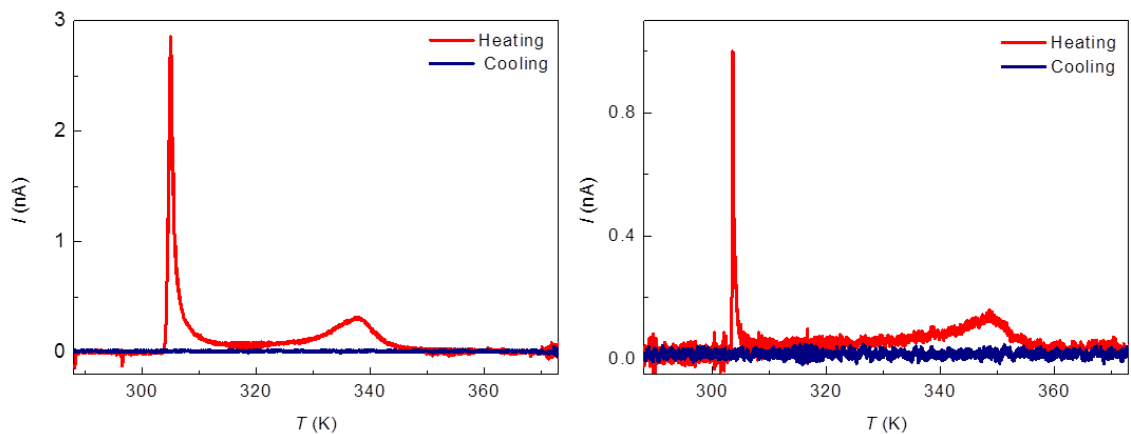

**Figure S10.** Electrical current measured at $V = 1$ V between 288 K and 330 K. Two resonances in current show up at roughly the same temperature than in the crystal in the main text.

Figure S11 shows the current measured as a function of increasing temperature in the same **1**·2CH$_3$CN crystal shown in the main text. The black line corresponds to the first thermal cycle on the sample. The resonance disappears in subsequent thermal cycles since the loss of acetonitrile is irreversible as explained in the main text. Strikingly, the resonance reappears at roughly the same temperature after the crystal has been exposed to liquid acetonitrile at room temperature.

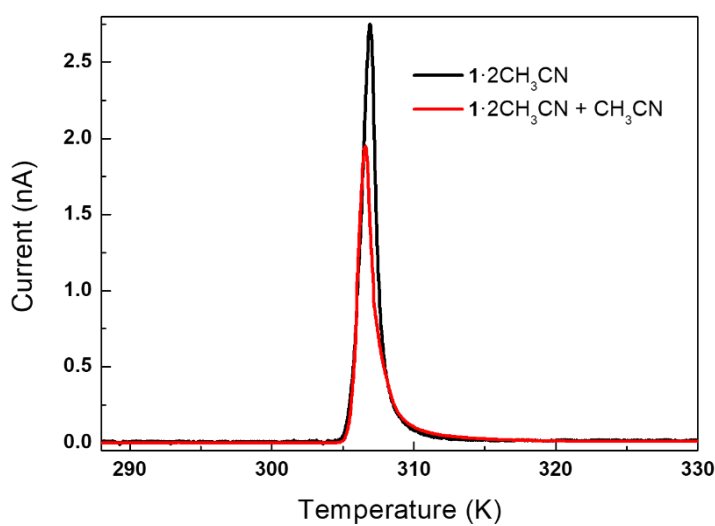

**Figure S11.** Electrical conductivity measurements between 288 K and 330 K. The black line corresponds to the original **1**·2CH$_3$CN (first thermal cycle) and the red one to the same crystal after a drop of acetonitrile was added in a subsequent thermal cycle.

Figure S12 shows an additional proof-of-concept acetonitrile sensor measurement. The current $I$ is measured across a **1**·2CH$_3$CN crystal while cycling the temperature between 288 K and 318 K. As shown in the main paper, a resonance in the current appears at a well-defined temperature only in those cycles where the crystal has been exposed to acetonitrile vapor.

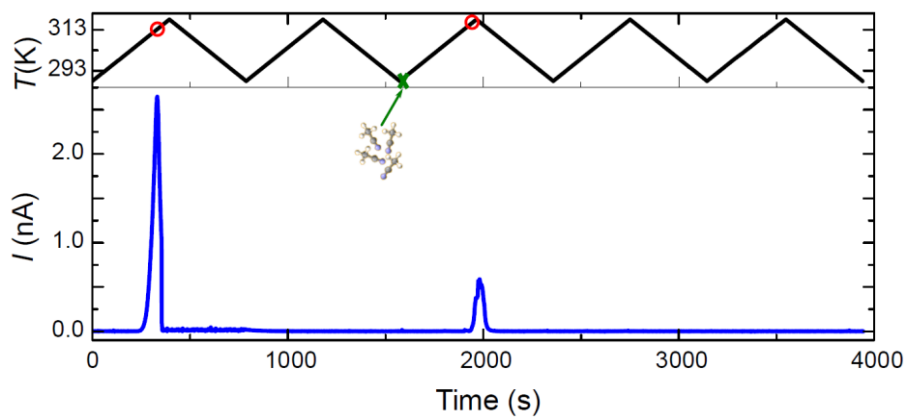

**Figure S12.** Current $I$ trace measured across a **1**·2CH$_3$CN crystal while cycling the temperature between 288 and 318 K. A resonance in the current (red circle) appears in the cycle where the **1**·2CH$_3$CN is exposed to acetonitrile vapor (green crosses).

## 8) Magnetic characterization of 1·2CH$_3$CN crystals

The magnetic susceptibility ($\chi_M$) is measured in a set of **1**·2CH$_3$CN crystals between 10 K and 380 K. The high-temperature $\chi_M T$ product is represented in Figure S13. The saturation value remains constant at 3.04 cm$^3$·K·mol$^{-1}$ from low temperatures up to 310 K. This value is characteristic of non-interacting high-spin iron (II) ions[1,2]. A small kink in the slope can be observed at 312 K and thereafter $\chi_M T$ decreases to 3.02 cm$^3$·K·mol$^{-1}$. The transition temperature matches with the first structural, optical and conductance transitions observed in the **1**·2CH$_3$CN crystal. According to Curie's law for a paramagnet, a change in $\chi_M T$ could be induced by a change in the spin value $S$ or a change in the $g$ factor. A high-spin to low-spin switch would have a more dramatic consequence in $\chi_M T$. The most probable scenario is therefore a change in $g$ due to a structural change around the Fe, as suggested by the DSC measurements and the DFT calculations.

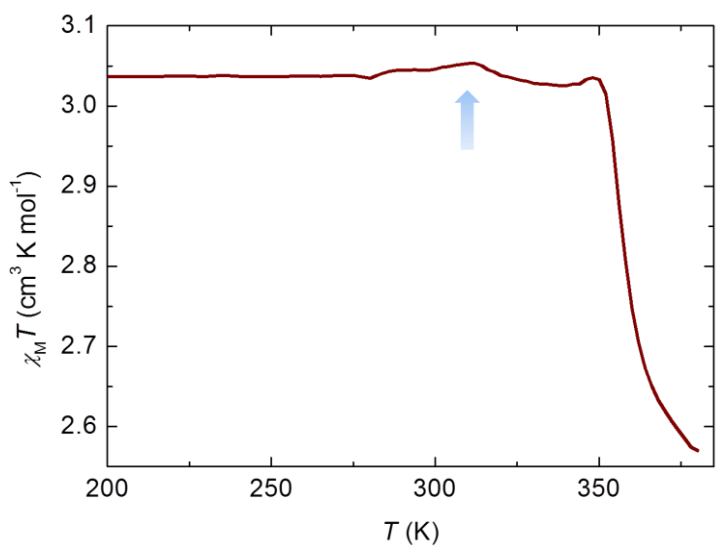

**Figure S13.** High temperature $\chi_M T$ measured as a function of the temperature in a set of **1**·2CH$_3$CN crystals. A small kink and a large drop in $\chi_M T$ are observed at 310 K and 350 K respectively.

After the first transition, an abrupt change in the slope shows up at 350 K; the second transition temperature in the conductance and the OR. The $\chi_M T$ value decreases down to 2.57 cm$^3$ K mol$^{-1}$ at the highest measured temperature. Note that the limitations in the set-up do not allow measuring above this temperature and therefore is difficult to determine whether $\chi_M T$ continues dropping to zero or stabilizes at a finite value. A sharp drop of $\chi_M T$ in these kind of materials is typically associated to a spin crossover transition to a low-spin state ($S = 0$) [1]. However, this transition typically occurs while decreasing the temperature and the bond lengths around the Fe ions decrease. In **1**·2CH$_3$CN crystals, such decrease in the bond length could be induced by the structural distortions caused by the release of acetonitrile, as suggested by DFT calculation (see below). The in-depth analysis of this effect is an interesting topic in itself that will be subject of a further study.

Figure S14 shows $\chi_M T$ measured in a **1**·2CH$_3$CN crystal at low temperatures. The value of $\chi_M T$ remains approximately constant down to 25 K. The sharp drop in $\chi_M T$ below 25 K may be indicative of antiferromagnetic correlations in the chains or the depopulation of excited magnetic levels.

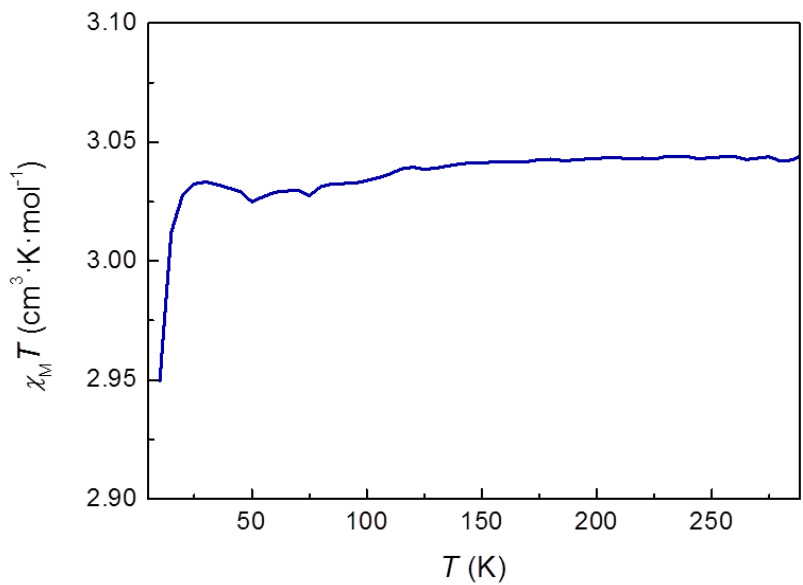

**Figure S14.** Low temperature $\chi_M T$ measured on a set of **1**·2CH$_3$CN crystals.

## 9)    DFT modelling.

DFT calculations are performed using PWSCF utility of QUANTUM ESPRESSO [3]. Exchange and correlation is treated using the PBE (Perdew-Burke-Ernzerhof) functional and a *Hubbard U* correction of 5 eV applied to d electrons in employed. Norm-conserving Rappe Rabe Kaxiras Joannopoulos pseudopotentials are used, and we set wave-function with charge density cutoffs of 60 Ry and 600 Ry, respectively. Geometrical optimizations are performed for **1** and **1**·2CH$_3$CN compounds until the forces on atoms are less than 0.003 eV/Å and the stress is less than 0.015 kbar. A unit cell containing 204 and 156 atoms is used for **1** and **1**·2CH$_3$CN, respectively. This implies a ferromagnetic alignment of the Fe ions along the chains (x direction). A Monkhorst-Pack grid of 5×1×1 is used for geometrical optimization while a 7×3×1 is employed for calculation of the density of states. After removal of interstitial CH$_3$CN molecules, the *a*, *b*, and *c* lattice parameters are shortened by 0.8 %, 1.8 % and 5.5%, respectively. For the **1**·2CH$_3$CN complex, we compute the following bond lengths around the each Fe ion: Fe-O=2.14 Å, Fe-N(CH$_3$CN)=2.18 Å and Fe-N(pz)=2.30 Å. For **1** we compute: Fe-O=2.18 Å, Fe-N(CH$_3$CN)=2.15 Å and Fe-N(pz)=2.27 Å. We compute a binding energy of 65 kJ/mol for acetonitrile.

The projected density of states shown in Figure S15 for **1**·2CH$_3$CN shows a large gap which is negligibly affected by removal of acetonitrile, consistent with states located far from the conduction and valence bands.

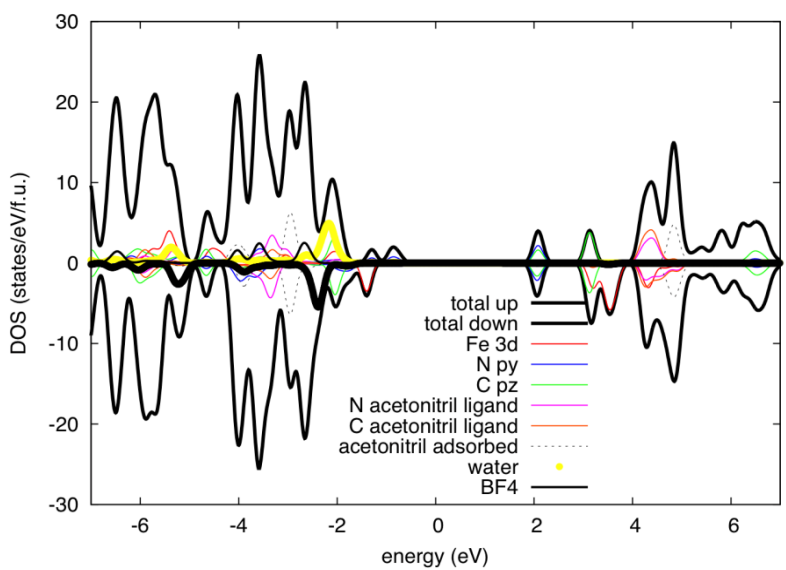

**Figure S15.** 1·2CH$_3$CN DFT calculations.

## References

[1] A. Białońska, R. Bronisz, K. Darowska, K. Drabent, J. Kusz, M. Siczek, M. Weselski, M. Zubko, A. Ozarowski, *Inorg. Chem.* **2010**, *49*, 11267–11269.

[2] J. S. Costa, S. Rodríguez-Jiménez, G. A. Craig, B. Barth, C. M. Beavers, S. J. Teat, G. Aromí, *J. Am. Chem. Soc.* **2014**, *136*, 3869–3874.

[3] P. Giannozzi, S. Baroni, N. Bonini, M. Calandra, R. Car, C. Cavazzoni, D. Ceresoli, G. L. Chiarotti, M. Cococcioni, I. Dabo et al., *J. Phys.: Condens. Matter* **2009**, *21*, 395502.